\UseRawInputEncoding
\documentclass[
 reprint,
 amsmath,amssymb,
 aps,
 prl,
]{revtex4-2}
\usepackage[normalem]{ulem}

\usepackage[utf8]{inputenc}
\usepackage{float}
\usepackage[normalem]{ulem}
\usepackage{graphicx}
\usepackage{dcolumn}
\usepackage{bm}
\usepackage{xcolor}
\usepackage{hyperref}
\usepackage{subfiles} 
\usepackage{booktabs}
\usepackage[normalem]{ulem}
\usepackage{newunicodechar}
\newunicodechar{−}{-}

\begin{document}

\preprint{APS/123-QED}

\title{
Entropy-production cost limits hierarchical decision-making in homing pigeon flocks}%

\author{Bibaswan Ghosh$^{1}$}
\thanks{Equal contribution}
\author{Vipul Rai$^{2}$}
\thanks{Equal contribution}
\author{Harsh Soni$^{2}$}
\author{Nitin Kumar$^{1}$}
\email{nkumar@iitb.ac.in}

\affiliation{
	$^1$Department of Physics, Indian Institute of Technology Bombay Powai, Mumbai 400076, India \\
	$^2$School of Physical Sciences, Indian Institute of Technology Mandi, Mandi 175005, India \\
	}
\date{\today}

\begin{abstract}
Collective decision-making requires individuals to reconcile diverse, often conflicting opinions to reach a common course of action. However, within a group, individuals rarely exert equal influence, leading to non-reciprocal communication patterns manifested as leader-follower relationships and social hierarchies. Here, using homing pigeon flocks as a model system, we develop a trajectory-based framework to infer and quantify hierarchy within navigating living collectives. We find that while small flocks exhibit pronounced hierarchy, increasing flock size progressively weakens it, driving the collective toward more egalitarian decision-making. We provide a physical rationale for this observation using a thermodynamic framework in which non-reciprocal communication between birds carries an information-entropy cost. Our calculations reveal that hierarchy weakening observed in larger flocks is consistent with maintaining a constant per-individual entropy production rate across flock sizes. Together, these results suggest that constraints on information-processing capacity may play an important role in the collective dynamics of living organisms.

\end{abstract}

\maketitle

\subsection{Main}
How interacting agents, both living and artificial, resolve conflicting preferences and reach a consensus while pursuing a common goal is an open question across disciplines \cite{axelrod1981evolution, carandini2013probing, mann2018collective, mcfarland1977decision, maroto2023systematic}. Achieving this often gives rise to social hierarchies, manifested through asymmetric leader–follower relationships and unequal influence among group members \cite{cheng2026asymmetric, king2009origins, couzin2005effective, nagy2010hierarchical}. In most cases, the underlying processes of communication and negotiation occur in an abstract decision-making space that is inaccessible to direct physical observation. However, goal-directed collective navigation phenomena in animals, such as homing and migration, provide an ideal platform for investigating these questions \cite{mandal2018animals, papi2012animal, fudickar2021animal, kaur2024prey, hays2020open}. This stems from the fact that individuals continuously adjust their motion towards a common destination, with collective decisions encoded in their spatial trajectories, providing a time-resolved record.

Among these systems, homing pigeon flocks are the most extensively studied experimentally \cite{bonadonna1997pigeon, wiltschko2007homing, dell2008flock, sasaki2022empirical}. When flying alone, individual pigeons exhibit distinct preferences for their homing routes, with varying degrees of route fidelity \cite{meade2005homing, sasaki2022empirical}. However, when released as a group, these individual preferences conflict, requiring the birds to dynamically resolve their navigational differences during flight, leading to a common trajectory towards home. Past studies using high-resolution GPS tracking data have revealed a strong hierarchical leader–follower organization within pigeon flocks \cite{biro2006compromise, nagy2010hierarchical, flack2013robustness, sasaki2022empirical}. Subsequent studies have investigated the physical traits that distinguish leaders from followers \cite{nagy2013context, watts2016lack, flack2012leaders}. Furthermore, these hierarchies are not static, as leadership shifts between individuals over the course of a flight \cite{watts2016misinformed, chen2016switching, zhang2014route}. But the fact remains that inter-bird communication during collective navigation is highly asymmetrical, breaking action–reaction and time-reversal symmetry. Therefore, analyzing collective decision-making using tools of active matter \cite{ramaswamy2010mechanics, osat2023non, fruchart2021non, tan2022odd,kumar2026active} and stochastic thermodynamics \cite{seifert2012stochastic, schnakenberg1976network, andrieux2004fluctuation} may provide valuable insights into the underlying phenomenon.

In this study, we are motivated by experimental results showing that the efficiency of homing or migrating birds increases with flock size \cite{dell2008flock, sasaki2022empirical, morford2026wisdom} and aim to understand the physical origins of this phenomenon. A simple wisdom-of-crowds-based explanation, called the Many-Wrongs hypothesis (MWH) \cite{bergman1964analysis, wallraff1978social, simons2004many}, in which collective error is evaluated by averaging individual flight routes with equal weights, is often used to explain this finding. However, T. Sasaki et al. \cite{sasaki2022empirical} discovered that MWH fails to account for quantitative aspects of the final flock trajectory. Therefore, we investigate whether the presence of an influence hierarchy within flocks can account for the observed homing paths. Addressing these requires a quantitative framework that can infer the strength of hierarchy directly from flight trajectories, while simultaneously characterizing the efficiency of the resulting navigation, which, as far as we know, does not exist. 

Here, using the trajectory dataset from reference \cite{sasaki2022empirical} as the empirical basis, we develop an integrated analytical and numerical framework to quantify hierarchical decision-making in homing pigeon flocks. We first formulate an analytical model of homing and then apply it to empirical trajectory data to extract a behavioral trait describing a flock's response to navigational errors. Then we construct a data-driven numerical model that reproduces the statistics of solo homing flights. Later, we introduce hierarchical decision-making by considering $N$ communicating agents whose individual opinions, represented by their navigational errors, are weighted according to their relative influence. The influence strengths are exponentially distributed across flock members with rate $\eta$, which we term the hierarchy index. Comparison with empirical flock trajectories reveals a systematic decrease in $\eta$ as flock size $N$ increases, indicating a progressive transition from hierarchical to more egalitarian decision-making. We also validate the globally inferred $\eta$ against fine-grained spatiotemporal dynamics by analyzing time delays associated with directional changes between pairs of birds. Finally, we show that the decreasing hierarchy with increasing flock size may originate from rapidly increasing information-entropy production rate per bird due to non-reciprocal, leader-follower communication.

\subsection{Analytical Homing Model}

\begin{figure}[t]
\includegraphics[width= 0.480\textwidth]{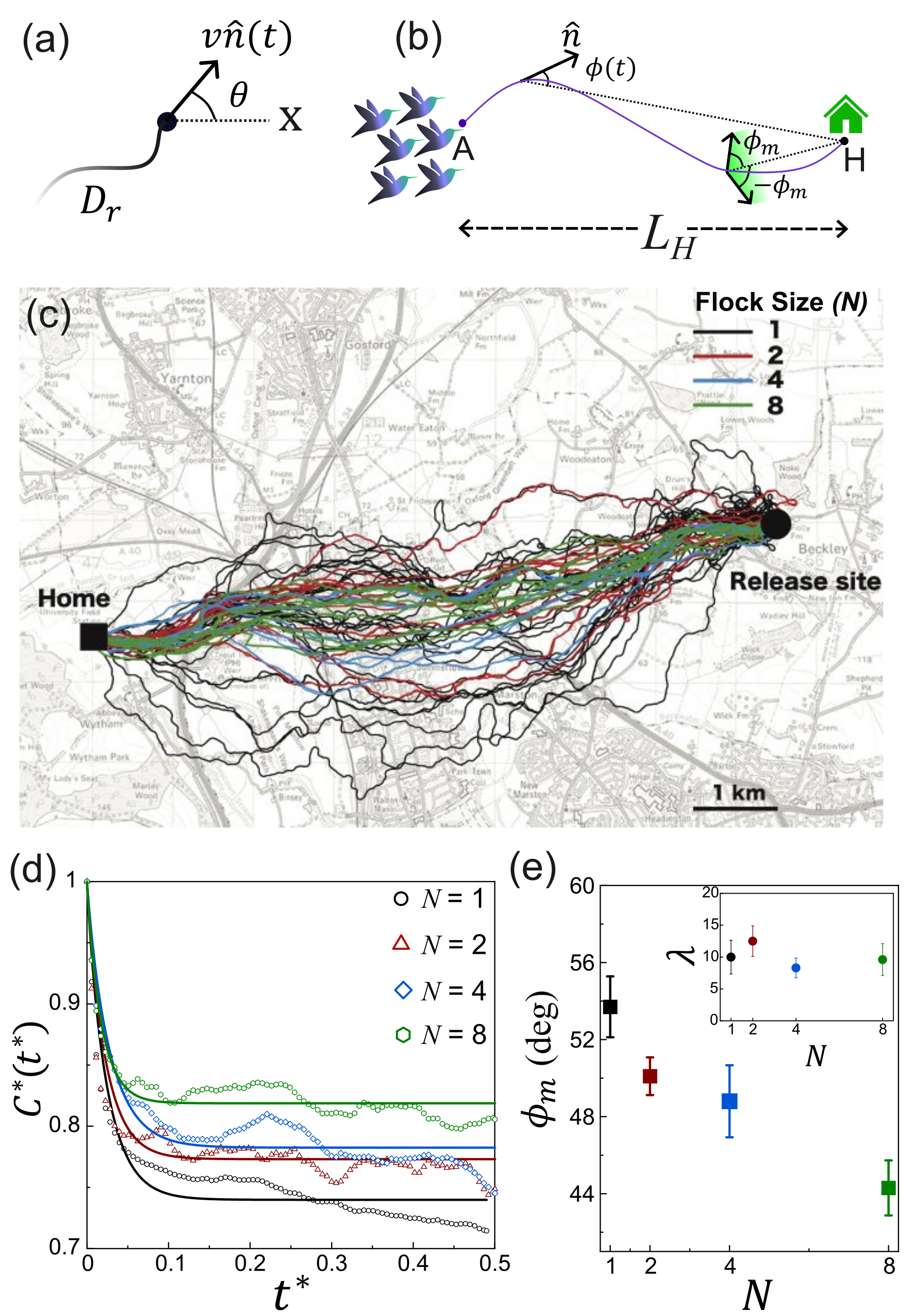}
\caption{\textbf{(a)} Schematic of a two-dimensional active Brownian particle (ABP) moving at a constant speed $v$, with its orientation angle ($\theta$) undergoing a random walk with diffusion constant $D_r$. \textbf{(b)} A pigeon flock modeled as an ABP, with its angular deviation from the home direction constrained within $\pm\phi_m$, facilitating homing. \textbf{(c)} Empirical data from ref. \cite{sasaki2022empirical} showing homing trajectories become progressively straighter and more efficient with increasing flock size. Image adapted from \cite{sasaki2022empirical}. \textbf{(d)} The dimensionless velocity autocorrelation function exhibits an increasing saturation plateau with flock size, reflecting enhanced homing efficiency. Solid lines represent fits to Eq. 1. \textbf{(e)} Fitted parameters $\lambda$ and $\phi_m$ as functions of flock size $N$. While $\lambda$ remains unchanged, $\phi_m$ decreases systematically with increasing $N$. Error bars equal standard deviation (SD).}
\label{fig1}
\end{figure}

We begin by formulating a generalized version of our earlier introduced minimal statistical model of homing dynamics \cite{paramanick2024uncovering, paramanick2024uncovering_erratum} (see Appendix A for detailed analytical calculation). We use the fact that the final homing path emerges from two simple components: (i) the animal's intrinsic stochasticity that causes deviations from the direct, straight line path to their destination; (ii) active course-corrections to prevent these deviations from accumulating, thus avoiding unsuccessful homing. Inspired by the active-matter paradigm we mimic the motion of an organism (or their collection as a single unit) as an overdamped active Brownian particle (ABP) \cite{solon2015active, paramanick2024programming} in two dimensions. It moves at a constant speed $v$ with its instantaneous orientation $\theta$ subject to angular noise with rotational diffusion constant value $D_r$ (Fig. \ref{fig1}\textbf{a}). To model homing, the ABP particle starts from $A$ and is set to reach home ($H$) with distance $AH = L_H$. We define $\phi(t)$ as the angle between the instantaneous direction of motion, $\hat{\mathbf{n}}$, and the line joining its current position to its home (dotted line in Fig. \ref{fig1}\textbf{b}). To incorporate the animal's response to navigational deviations, we define an angle $\phi_m$ such that $\phi(t)$ remains restricted between $+\phi_m$ and $-\phi_m$ leading to a directed motion towards $H$. Thus, $\phi_m$ quantifies the animal's tolerance to navigational errors, providing a measure of its behavioral response during homing.

To quantify the statistical properties of the resulting trajectories, we use the temporal velocity autocorrelation function averaged over multiple trajectories: $C(t)=\left\langle \hat{\mathbf n}(t'+t)\cdot\hat{\mathbf n}(t')\right\rangle_{t'}$, where angular brackets denote average over the initial time $t'$ spanning the entire duration of the trajectory (Eq. 32 in Appendix A). We introduce the dimensionless time $t^{*}=vt/L_H$ and the dimensionless stochasticity parameter $\lambda=L_H/L_P=L_HD_r/v$ where $L_P=v/D_r$ is the persistence length of the ABP. In terms of these variables, the model yields the following analytical expression for the dimensionless velocity autocorrelation function.

\begin{equation}
\begin{aligned}
C^{*}(t^*)
=&\frac{\sin^2\phi_m}{\phi_m^2}
+\frac{2}{\phi_m^2}
\Big[\sin^2\phi_m\sum_{n=1}^{\infty}
\frac{
\exp\!\left(-\lambda\xi^2 t^*\right)
}
{
\left(1-\xi^2\right)^2
}
\\
&+
\cos^2\phi_m
\sum_{n=1}^{\infty}
\frac{
\exp\!\left(-\lambda\zeta^2 t^*\right)}
{
\left(1-\zeta^2\right)^2
}\Big]
\end{aligned}
\label{HE}
\end{equation}

Here, $\xi = n\pi/\phi_m$ and $\zeta = (2n-1)\pi/2\phi_m$. Clearly, at short times, the two summation terms produce a rapid decay, reflecting the stochastic angular fluctuations of the trajectory relative to the straight line. This is quantified by the parameter $\lambda$. At long times, the correlation approaches a finite saturation plateau, $\sin^2\phi_m/\phi_m^2$, which arises from the bounded angular deviation imposed by the homeward direction. Thus, evaluating the $C^*$ from GPS coordinates of trajectories and comparing it with Eq. \ref{HE} provides a simple procedure for independently extracting two fundamental behavioral parameters of homing: the path stochasticity, $\lambda$, and the tolerance to navigational error, $\phi_m$. We successfully validate this expression on empirical trajectories from two distinct animal species (SI section I and Fig. S1), highlighting its universality and generality as a minimal model.

Next, we apply this model to empirical trajectories of homing pigeon flocks of different sizes, $N$, using the GPS data provided in reference \cite{sasaki2022empirical} (see the SI section II for more details). The original experiments involved 30 pigeons, which were released toward their home loft traveling a distance of $L_H=8.1$ km in the randomly mixed flock sizes of $N=1,2,4,$ and $8$ (further details available in Method details section of \cite{sasaki2022empirical}). As shown in Fig. \ref{fig1}\textbf{c}, individual pigeons exhibit tortuous homing routes with wide variation among individuals, which become systematically straighter and more efficient as $N$ increases. For each $N$, we calculate the $C^{*}(t^{*})$ (SI section II) using their GPS trajectory data as shown in Fig. \ref{fig1}\textbf{d}. Solid lines represent fits with Eq. \ref{HE}, which we use to extract parameters $\lambda$ and $\phi_m$ (Fig. \ref{fig1}\textbf{e} ). Interestingly, while the intrinsic stochasticity ($\lambda$) remains essentially unchanged across flock sizes (see inset of Fig. \ref{fig1}\textbf{e}), $\phi_m$ decreases systematically with $N$ . This suggests that the improvement in collective homing arises solely from a systematic decrease in the birds' tolerance for navigational errors. 

\subsection{Simulating communicating homing pigeons}

\begin{figure}[t]
\includegraphics[width= 0.480\textwidth]{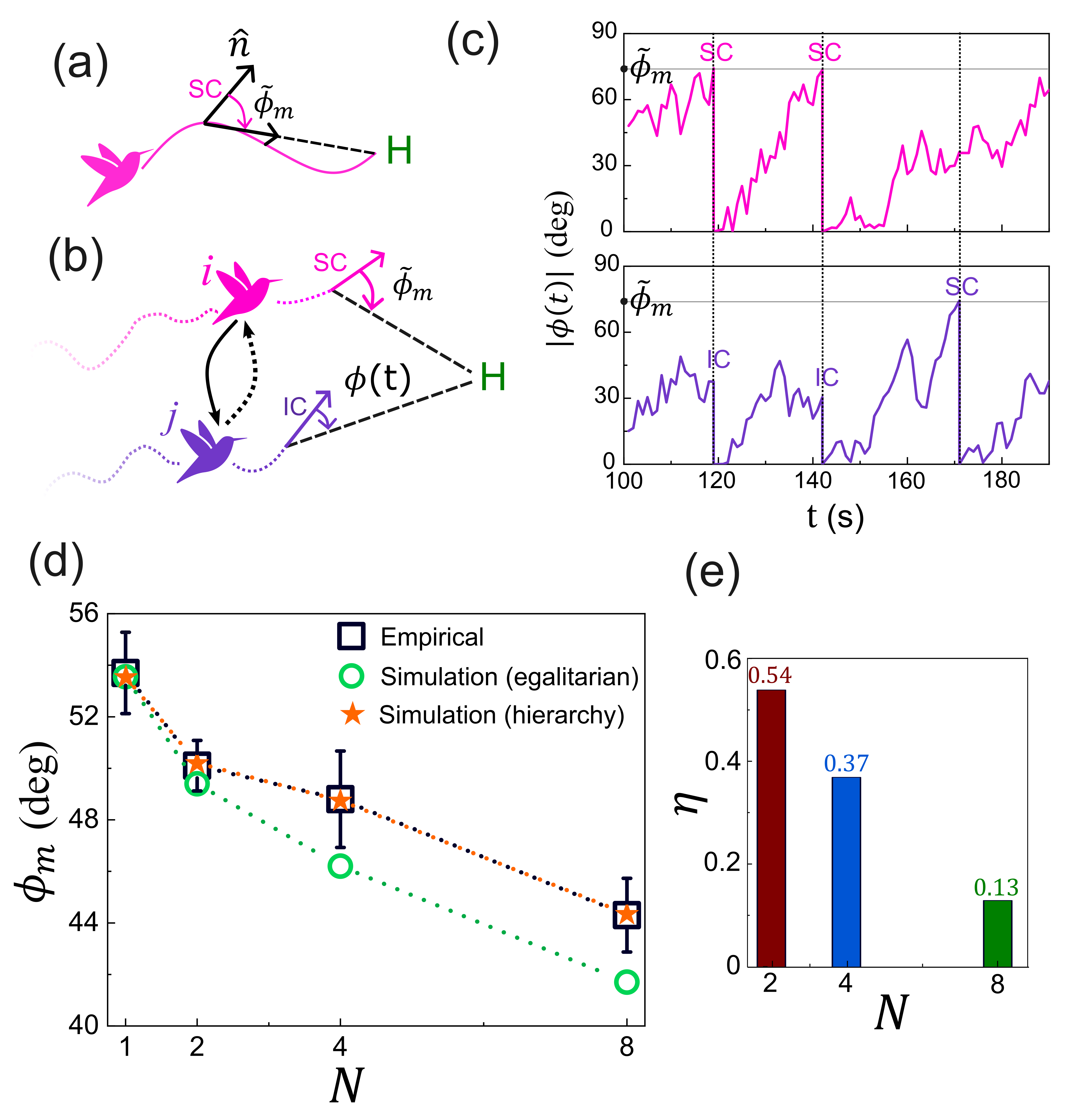}
\caption{\textbf{(a)} Each individual agent undergoes a self-correction (SC) event, resetting its angular deviation from $\pm\tilde{\phi}_m$ to zero. \textbf{(b)} In a flock, an SC event in agent $i$ can also trigger an induced correction (IC) event in agent $j$, resetting its angular deviation from $\pm\phi(t)$ to zero even when $|\phi(t)|<|\tilde{\phi}_m|$. Solid and dashed curved arrows between birds represent non-reciprocal interactions. \textbf{(c)} Schematic of non-reciprocal communication, in which an SC event in one agent induces an IC event in the other with unequal probabilities ($P_i \ne P_j$). \textbf{(d)} Egalitarian simulations ($P_i=1$ for all $i$) fail to capture the observed dependence of $\phi_m$ on flock size $N$, whereas the trend is recovered by introducing an appropriate degree of hierarchy for each $N$. Error bars equal SD. \textbf{(e)} The corresponding hierarchy index $\eta$ required to reproduce the $\phi_m$ values in \textbf{(d)}, showing a systematic decrease with increasing $N$.}
\label{fig2}
\end{figure}

To this end, we develop a data-driven agent-based simulation of interacting homing pigeons. Our first step is to replicate a statistically equivalent trajectory for a solo pigeon in a computer simulation (SI section IIIA). By fitting Eq. \ref{HE} to solo trajectory data, we extract parameters $\lambda = 10$ and $\phi_m = 54^{\circ}$. Using $v$ = 55 km h$^{-1}$ and $L_H=8.1$ km, we find the $D_r = v\lambda/L_H = 0.02$ rad$^2$s$^{-1}$ of our numerical ABP agent. Note that our analytical model assumes a bounded angular deviation within $[-\phi_m,\phi_m]$ assuming a uniform stationary distribution of $\phi$ between these two extremes. However, the empirical $\phi$ distribution for solo pigeons is approximately Gaussian, centered at zero (see SI Fig. S2\textbf{c}). To account for observed $\phi$ statistics without altering $C^{*}(t^*)$ and $\phi_m$ values, we modify the protocol in the numerical implementation. Specifically, we define a threshold angle $\tilde\phi_m$, beyond which (i.e. $\phi(t)>\tilde\phi_m$ and $\phi(t)< -\tilde\phi_m$) the agent's orientation is instantaneously reset to $\pm\tilde{\phi_r}$ respectively (see SI Fig. S2\textbf{a}), a mechanism analogous to stochastic resetting \cite{evans2020stochastic, kumar2020active, biswas2025target, barman2025optimizing}. We find that among many possible combinations of $\tilde\phi_m$ and $\tilde\phi_r$, their respective values of $74^{\circ}$ and $0^{\circ}$ reproduce the empirical distribution of $\phi_m$ most accurately (SI section IIIB and SI Fig. S2\textbf{c}). In conclusion, the final homing scheme of a single agent is as follows: Whenever $|\phi(t)|$ exceeds $|\tilde\phi_m|$, the agent undergoes an instantaneous course correction, resetting its heading $\phi\rightarrow 0^\circ$, i.e., facing directly towards home. We refer to this as a self-correction (SC) event (Fig. \ref{fig2}\textbf{a}).

Now we extend the numerical model to a group of $N$ identical, communicating agents that begin homing simultaneously. Each agent follows the same SC protocol described above. Agents are coupled through social communication, i.e., each SC event broadcasts an instantaneous global signal to all other flock members. Upon receiving this signal from agent $i$, every other agent $j$ may reorient toward home regardless of whether its own angular deviation satisfies $|\phi(t)| \geq |\tilde\phi_m|$ (Fig. \ref{fig2}\textbf{b} $\&$ \textbf{c}). We refer to this as an induced correction (IC). In this way, one agent's decision to correct its own course can simultaneously trigger a directional correction across the rest of the flock, representing a minimal mechanism for collective decision-making through inter-agent communication.

We start with egalitarian (reciprocal) communication, where all individuals exert equal social influence. In this limit, whenever any agent undergoes an SC event, every other member of the flock also undergoes an IC event, which is also equivalent to the traditional MWH. Averaged over 1000 simulation trajectories for all flock sizes, the egalitarian protocol reproduces the decrease in $\phi_m$ with increasing $N$ only qualitatively. As shown in (Fig. \ref{fig2}\textbf{d}, circle symbols), the predicted error-tolerance cone is considerably narrower than that observed in real flocks. Thus, fully egalitarian decision-making fails to reproduce experimental results. 

To account for this discrepancy, we introduce social hierarchy by assigning unequal influence weights to each agent. Specifically, each agent is assigned a rank-dependent influence probability $P_i = e^{-\eta(i-1)}$, where $i$ is its hierarchical rank and $\eta$ ($>0$) is a hierarchy index that controls the degree of influence asymmetry within the group. Here, the highest-ranked individual is assigned $P_1 = 1$. In real pigeons, studies have suggested that leaders are determined by factors such as an individual's route fidelity \cite{meade2005homing, sasaki2022empirical} or their flight speed \cite{pettit2015speed} during solo flights. In our simulation, it is irrelevant, as all agents are equivalent. In this new protocol, when an agent of rank $i$ undergoes an SC event, the other $N-1$ agents undergo IC with probability $P_i$. The egalitarian case is recovered for $\eta = 0$, where $P_i = 1$ for all $i$. At the opposite extreme, $\eta \rightarrow \infty$ reduces to a single-leader limit where only the top-ranked bird retains finite influence ($P_1 = 1$, $P_{i>1} \rightarrow 0$), and collective corrections are driven entirely by one leader. Next, we systematically vary $\eta$ for different $N$ and identify the value that best reproduces the mean value of experimentally measured $\phi_m$ (Fig. \ref{fig2}\textbf{d} star symbol). Interestingly, we find that the inferred $\eta$ decreases systematically with $N$ (Fig. \ref{fig2}\textbf{e}), revealing a size-dependent transition from strongly hierarchical decision-making in small groups towards increasingly egalitarian decision-making in larger flocks.


\subsection{Validating hierarchy index on directional delays between pigeons}

Due to its minimal nature, our framework provides only a coarse-grained description of the system and omits detailed microscopic interactions among individual agents. Moreover, all error-correction events in numerical simulations are instantaneous. In reality, navigating birds have finite sensory and cognitive reaction times. Consequently, an induced error-correction events between birds must have a finite time delay. If our hierarchy-based communication model provides a valid description of the collective decision-making process, the systematic decrease of the inferred $\eta$ with $N$ should also leave a signature in the short-time dynamics of the birds. We therefore turn to high-resolution spatiotemporal trajectory data to examine the finite delays in the propagation of directional information between individuals.

\begin{figure*}[t]
\centering\includegraphics[width= 0.964\textwidth]{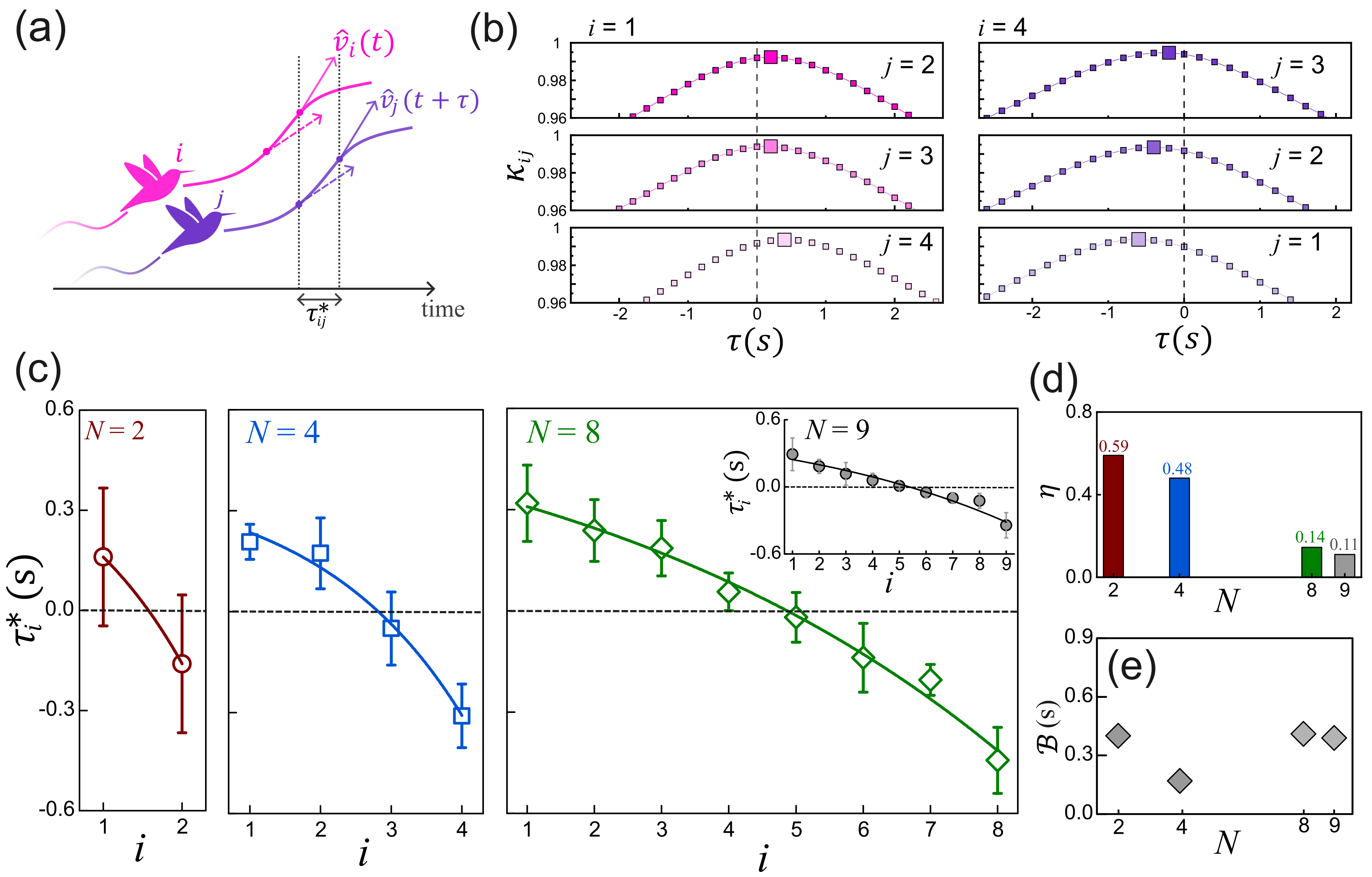}
\caption{\textbf{(a)} Schematic illustrating how bird $i$ induces a directional change in bird $j$ after a characteristic time delay $\tau^*_{ij}$. \textbf{(b)} Representative pairwise directional cross-correlation functions, $\kappa_{ij}(\tau)$. For a leader ($i=1$), $\kappa_{ij}(\tau)$ peaks at positive time delays for all other birds $j$, with enlarged symbols marking the corresponding $\tau^*_{ij}$ values. Similar plots for a follower ($i=4$) show negative $\tau^*_{ij}$ values for all other birds. \textbf{(c)} Average delay times, $\tau^*_i$, for individual agents across different flock sizes $N$. Solid lines represent fits to Eq. \ref{eq:delay_fit} while treating $\eta$ and $\mathcal B$ as free parameters. Inset: corresponding results for the experimental data from Ref. \cite{nagy2010hierarchical}. Error bars equal SD. \textbf{(d)} The $\eta$ values showing similar trend as in Fig. \ref{fig2}\textbf{e}. \textbf{(e)} The parameter $\mathcal B$ remains largely independent of the flock size.}
\label{fig3}
\end{figure*}

We start by computing the pairwise directional cross-correlation function between all individuals within a flock, $\kappa_{ij}(\tau) = \langle \hat{\mathbf{v}}_{i}(t) \cdot \hat{\mathbf{v}}_{j}(t+\tau) \rangle$, where $\hat{\mathbf{v}}_i$ and $\hat{\mathbf{v}}_j$ are the unit velocity vectors of birds $i$ and $j$ at time and $\tau$ denotes the temporal lag (see schematic Fig. \ref{fig3}\textbf{a}). Following the approach of Nagy et al. \cite{nagy2010hierarchical}, we define the characteristic delay time between the two individuals as the lag that maximizes the cross-correlation, $\tau^*_{ij} = \underset{\tau}{\operatorname{arg\,max}}\,
k_{ij}(\tau)$ (see SI section V). With this convention, a positive $\tau^*_{ij}$ indicates that directional changes of bird $i$ occur earlier than those of bird $j$ ($i = 1$ in Fig. \ref{fig3}\textbf{b}), whereas a negative value indicates that bird $j$ leads bird $i$ ($i = 4$ in Fig. \ref{fig3}\textbf{b}). We then average the pairwise delay times for each $i$ over all other flock members and across multiple flight trials, assigning an average lag time $\tau^*_i$ to each bird. This quantity provides a microscopic measure of relative influence rank: on average, individuals with positive $\tau^*_i$ tend to initiate directional changes, whereas those with negative values tend to follow the directional changes of others. We rank the birds within each flock in descending order of their $\tau^*_i$ values and plot as a function of hierarchical rank, $i=1,\ldots, N$, for $N=2$, $4$, and $8$  (solid symbols in Fig. \ref{fig3}\textbf{c}). Interestingly, the profiles remain bounded within nearly the same maximum and minimum time values across flock sizes. Moreover,  these times remain bounded between ($\approx \pm 0.25$ s), which is remarkably close to the reaction times reported in pigeons in other independent studies \cite{blough1978reaction, wang1992time}.

To capture the observed $\tau_i^*$ curves, we propose a phenomenological relationship: $\tau^*_i  = \mathcal{A}-\mathcal{B}/P_i$ where $\mathcal{A}$ and $\mathcal{B}$ are positive constants with dimensions of time. Because pairwise delays are antisymmetric ($\tau^*_{ij} = -\tau^*_{ji}$), $\tau^*_i$ satisfy $\sum_{\substack{i}}\tau^*_{i} = 0$ As a result, parameters $\mathcal A$ $\&$ $\mathcal B$ are related to each other: $\mathcal A = \mathcal B \langle1/P\rangle$ where $\langle1/P\rangle$ is the flock average of $1/P_i$. This renders only a single free timescale, $\mathcal B$, which should be closely related to characteristic physiological timescales associated with sensory response and aerodynamic inertia of each bird, and therefore is expected to remain independent of flock size $N$. Substituting the exponentially heterogeneous rank-dependent influence probability, $P_i=e^{-\eta(i-1)}$, introduced in our numerical model, yields a simple and testable relation

\begin{equation} \tau^*_i = \frac{\mathcal B(e^{N\eta}-1)}{N(e^\eta-1)} - \mathcal B e^{\eta(i-1)} \label{eq:delay_fit} \end{equation}

where the first term is evaluated from $\mathcal B \langle1/P\rangle =  \mathcal (B/N) \sum_{\substack{i}}e^{\eta(i-1)}$. 

We keep both $\mathcal B$ $\&$ $\mathcal \eta$ as free parameters and fit the data points shown in Fig. \ref{fig3}\textbf{c}. Clearly, fits represented as solid lines accurately predict the empirical trend. As a further test, we also analyze an independent dataset from reference \cite{nagy2010hierarchical}, which reports experiments with a flock of size $N=9$ (see SI section VI). Surprisingly, this dataset also fits Eq. \ref{eq:delay_fit} quite well. Moreover, the fitted values of $\eta$ (Fig. \ref{fig3}\textbf{d}) also match those obtained from homing simulation (Fig. \ref{fig2}\textbf{e}) reasonably well. Moreover, $\mathcal B$ remains approximately independent of $N$ (Fig. \ref{fig3}\textbf{e}), consistent its their interpretation as characteristic timescales. In conclusion, this analysis corroborates our earlier finding of increasingly reciprocal and egalitarian decision-making in homing pigeons as flock size increases.

\subsection{\texorpdfstring{Entropy production rate bound explains the observed $\eta$ vs. $N$ trend}{Entropy production rate bound explains the observed eta vs. N trend}}

\begin{figure}[t]
\centering\includegraphics[width= 0.45\textwidth]{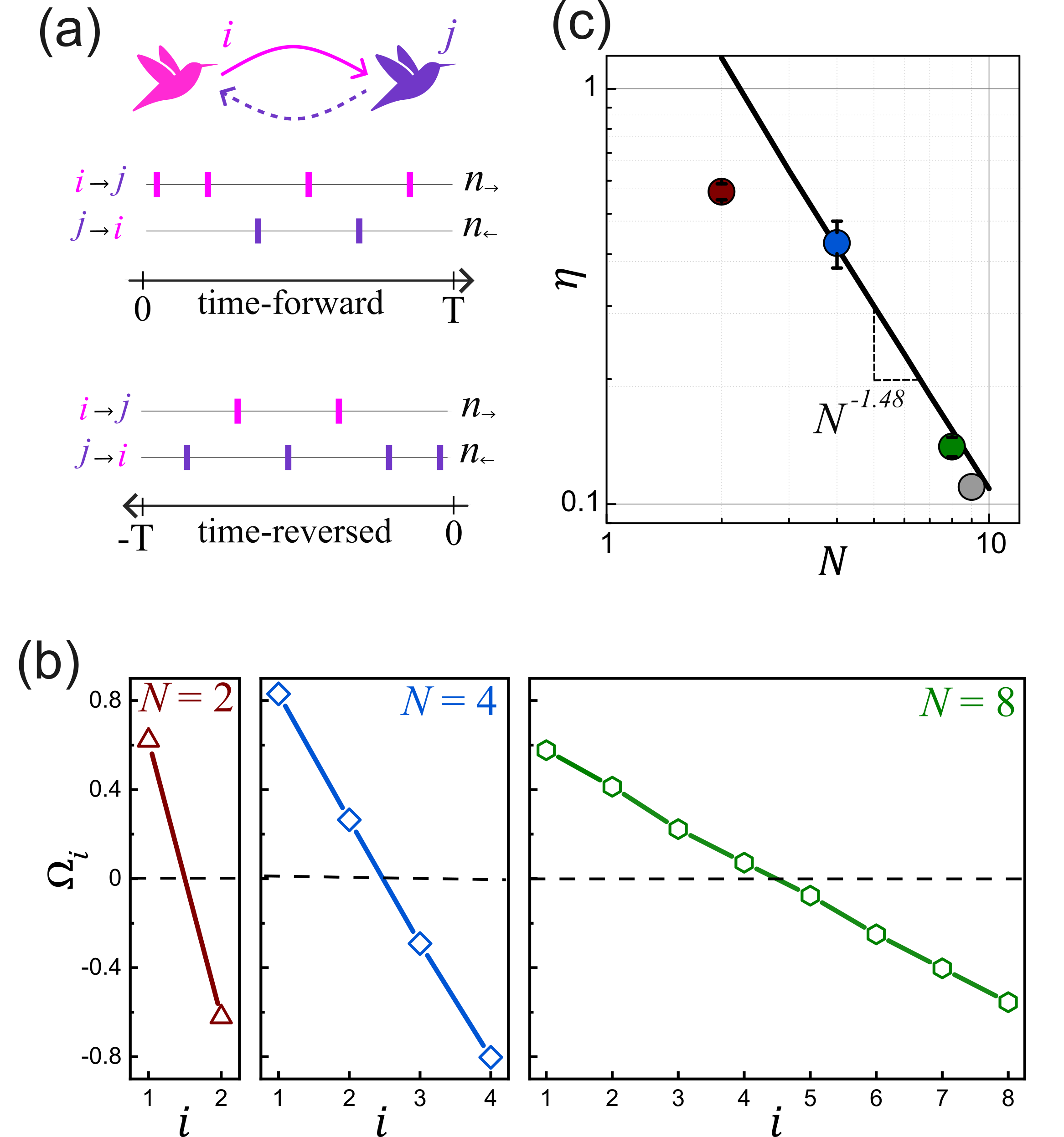}
\caption{\textbf{(a)} Top: A pair of birds $i$ $\&$ $j$, communicating asymmetrically with $i$ having greater influence than $j$. Bottom: A schematic representing the time-forward and time-reversed histories of IC events, with $n_{\rightarrow}$ ($n_{\leftarrow}$) the number of events from $i$ to $j$ (from $j$ to $i$); the time-reversed history is less probable. \textbf{(b)} Thermodynamic affinity per agent for three flock sizes, showing that both its maximum and minimum values remain bounded, similar to what was observed in their $\tau_i^*$ plots in Fig. \ref{fig3}\textbf{c}.  \textbf{(c)} Solid line shows the numerical fit of Eq.~\eqref{eq:final_result} keeping $\dot s$ constant ($ = 0.009$ $k_B$s$^{-1}$), successfully capturing the decreasing trend of $\eta$ with $N$.}
\label{fig4}
\end{figure}

Finally, we provide a plausible physical rationale for the observed decreasing trend of $\eta$ with $N$. We quantify the informational cost of non-reciprocal decision-making through the entropy production rate (EPR) of the
communication process \cite{loos2020irreversibility, Zhang_PRR_L022033,Ferretti_PhysRevE.106.034608}. Let $\mathcal{F}_{i\rightarrow j}$ ($\mathcal{F}_{j\rightarrow i}$) denote the number of successful IC events from $i$ to $j$ ($j$ to $i$) per unit time. Since each IC event in $j$ originates from an SC event in $i$, $\mathcal{F}_{i\rightarrow j} = \mathcal {R}_i P_i$ where $\mathcal {R}_i$ denotes the rate of SC events occurring in the $i^{th}$ agent during the entire homing process. Similarly, $\mathcal{F}_{j\rightarrow i} = \mathcal {R}_j P_j$. Treating them as independent Poisson processes~\cite{Kingman1992,VanKampen2007}, we assign $\mathcal P^+ $ as the probability density of a particular sequence of IC events between $ij$ pair (number of events $n_{\rightarrow}$ and $n_{\leftarrow}$ from $i \rightarrow j$ and $j \rightarrow i$ respectively over time $T$) in the time-forward direction as shown in [Fig.~\ref{fig4}\textbf{a}]. In its time-reversed counterpart, the direction of each event is reversed, and the corresponding probability density is given by $\mathcal P^{-}$. Due to non-reciprocal communication, $\mathcal P^+ \neq \mathcal P^-$, thus breaking the time-reversal symmetry. To measure a certain lower bound on the EPR ($\dot S_{ij}$), we evaluate the Kullback-Leibler divergence rate between the time-forward and time-reversed sequence of events (see Appendix B for detailed calculation), in units of $k_B$:  

\begin{equation}
\dot S_{ij} = \frac{1}{T}\Bigl\langle\ln\frac{\mathcal P^+}{\mathcal P^-}\Big\rangle=(\mathcal{F}_{i\rightarrow j}-\mathcal{F}_{j\rightarrow i})
\ln\frac{\mathcal F_{i\rightarrow j}}{\mathcal{F}_{j\rightarrow i}}
\end{equation}

Note that this is identical to the Schnakenberg EPR for a bidirectionally communicating pair as a product of thermodynamic flux (term in parentheses) and affinity $\Omega_{ij}\equiv\ln\left(\mathcal F_{i\rightarrow j}/ F_{j\rightarrow i}\right)$ ~\cite{schnakenberg1976network,
hinrichsen2013entropy}. Next, we define the mean affinity per connection for the $i^{\mathrm{th}}$ agent as $\Omega_i=[1/(N-1)]\sum_{\substack{j=1  \\ j\neq i}}^{N}\Omega_{ij}$, Here, the sign of $\Omega_i$ reveals the direction of the net information flow originating from agent $i$, with positive (negative) indicating an outward (inward) direction. Interestingly, the affinity profiles shown in Fig. \ref{fig4}\textbf{b} bear a close resemblance with $\tau^*_i$ plots of Fig. \ref{fig3}\textbf{c}.

Our simulations show that $\mathcal R_i \approx \mathcal R_j$ in a flock of given size (see SI Fig. S4\textbf{a}) which we approximate with $\mathcal R_0$. Since two agents are always communicating independently and substituting $P_i = e^{-\eta(i-1)}$ and $P_j = e^{-\eta(j-1)}$, we can calculate the total EPR by summing over all pairs for $N$ agents: 
\begin{equation}
\dot{S} = \frac{1}{2}\mathcal R_0
\sum_{i=1}^{N}
\sum_{\substack{j=1\\j\neq i}}^{N}
\eta(j-i)
\left[
e^{-\eta(i-1)}
-
e^{-\eta(j-1)}
\right].
\label{eq:initial_sum}
\end{equation}

The factor of half in front avoids double-counting of a pair. By evaluating the above summation (see Appendix C), and calculating the EPR per individual (i.e. $\dot s = \dot S/N$), we find:

\begin{equation}
\begin{aligned}
\dot{s} &=
\frac{\mathcal R_0 \eta}{2(1-e^{-\eta})^2}
\Big[
(N-1)-(N+1)e^{-\eta} \\
&\qquad +(N+1)e^{-N\eta}
-(N-1)e^{-(N+1)\eta}
\Big].
\end{aligned}
\label{eq:final_result}
\end{equation}
Encouragingly, in the limit $N\eta \ll 1$ which corresponds to small deviations from equilibrium, $\dot s\sim \mathcal{R}_0\eta^2 N^3$. The quadratic dependence on $\eta$ is consistent with Onsager's reciprocity relations for small deviations from equilibrium \cite{andrieux2004fluctuation}. Moreover, $\dot{s}$ vanishes as $\eta \rightarrow 0$, when communication becomes reciprocal. 

Next, to understand the observed decreasing trend of $\eta$ with $N$, we assume $\dot{s}$ remains independent of $N$. We use this condition to fit the empirically obtained average of the $\eta$ values in Fig. \ref{fig2}\textbf{e} and Fig. \ref{fig3}\textbf{d} for $N$ = 4, 8 $\&$ 9 ignoring $N = 2$. Further, we incorporate the simulation finding that $\mathcal R_0$ decays weakly with $N$ [$\mathcal R_0 = 0.026 N^{-0.25} \mathrm{s}^{-1}$, SI Fig. S4\textbf{b}]. As shown in Fig. \ref{fig4}\textbf{c}, this condition captures the general empirical trend of $\eta(N)$ reasonably well (fitted $\dot s = 0.009$ $k_B$s$^{-1}$), suggesting that the observed weakening of the hierarchy with increasing flock size may originate from preventing the EPR per individual from diverging as the flock size grows.


To conclude, we develop a multi-scale physical framework that connects coarse-grained homing performance, microscopic directional delays, and stochastic thermodynamics to study decision-making in animal collectives. Since the procedure only requires analyzing trajectory coordinates, it is highly versatile and robust. In this study, we apply this framework to high-resolution GPS trajectories of homing pigeon flocks and infer the influence hierarchy directly from observed flight data. The central finding is that small flocks exhibit hierarchical decision-making sustained by strong leader–follower asymmetries ($\eta$), whereas increasing flock size progressively weakens this asymmetry, driving the collective toward more reciprocal, egalitarian communication. Microscopic directional lag times provide an independent confirmation of this transition across flock sizes ($N$). We also provide a thermodynamic explanation of the phenomenon where hierarchy weakening emerges as a direct consequence of limiting the per-individual entropy production rate. Our calculations yield an $\eta$ vs. $N$ trend that is in close agreement with the experimental trend.

A few aspects of our results are worth highlighting. First, we assume bidirectional communication that ensures that each time-forward induced-correction event has a finite reverse probability and may not strictly hold for all pairs inside larger flocks. Nevertheless, as a minimal coarse-grained model, this approximation is justified. Secondly, previous studies on starling flocks suggest that each bird interacts with $6-7$ other birds topologically \cite{ballerini2008interaction}. Our results also provide a possible thermodynamic rationale for this finding, as maintaining non-reciprocal communication with many birds may incur a substantial entropy production rate (EPR). Third, non-reciprocal collective dynamics have also been observed in artificial active-matter systems \cite{lavergne2019group,Kant_064602,H_Soni}. Since our framework provides a simple and practical approach to inferring influence ranks and hierarchies from trajectory data, such systems, including more sophisticated programmable robots \cite{paramanick2024programming, fruchart2021non} could therefore serve as a useful testing ground for examining the broader applicability of the limiting-EPR hypothesis proposed here.

\textit{Acknowledgments} - N.K. acknowledges financial support from Anusandhan National Research Foundation (ANRF) for Advanced Research (Grant No. ANRF/ARG/2025/005689/PS) and IITB. NK also thanks Dibyendu Das and Sai Vinjanampathy for insightful discussions. HS acknowledges SERB for the SRG (grant no. SRG/2022/000061-G).

\bibliography{references}

\appendix

\section{APPENDIX A:
THEORETICAL MODEL OF TEMPORAL ORIENTATION
AUTOCORRELATION FUNCTION}

The homing dynamics are described by
\begin{equation}
\begin{aligned}
    \dot r &= -v\cos\phi, \\
    \dot\phi &= \frac{v}{r}\sin\phi+\dot\theta,
\end{aligned}
\end{equation}
where $r$ is the distance from home, $\phi$ is the heading
relative to the instantaneous home direction, and $\theta$
is the heading relative to the fixed $x$-axis.
For $r\gg v/D_r$, the home direction varies slowly on the
rotational diffusion timescale. Neglecting the geometrical
term therefore gives~\cite{paramanick2024uncovering}
\begin{equation}
    \dot\phi \simeq \dot\theta.
    \label{eq:large_distance}
\end{equation}
Since $\theta$ undergoes rotational diffusion with coefficient
$D_r$, Eq.~\eqref{eq:large_distance} implies that the
probability density $P(\phi,t)$ obeys
\begin{equation}
    \frac{\partial P}{\partial t}
    = D_r\frac{\partial^2 P}{\partial\phi^2}.
\end{equation}
and is confined to $[-\phi_m,\phi_m]$. With $P(\phi,0)=\delta(\phi)$ and
reflecting boundaries $\partial_\phi P(\pm\phi_m,t)=0$, which conserve
probability, we obtain

\begin{equation}
\begin{aligned}
P(\phi,t) = \frac{1}{2\phi_m} \Bigg[ 1 + 2 \sum_{n=1}^{\infty}
&\exp\!\left( -\left(\frac{n\pi}{\phi_m}\right)^2 D_r\,t \right) \\
&\times \cos\!\left(\frac{n\pi\,\phi}{\phi_m}\right) \Bigg]
\end{aligned}
\end{equation}

 So we are calculating sliding average, so angular displacement is $\Delta\phi(t) \equiv \phi(t'+t) - \phi(t)$. Then correlation is :
\begin{equation}
C^{\text{*}}(t) = \left\langle \cos\!\big[\Delta\phi(t)\big] \right\rangle_{t'}
\end{equation}
Correlation requires two ingredients:
\begin{enumerate}
\item The \textbf{steady-state} distribution of the starting angle $\phi'=\phi(t')$:
\begin{equation}
P_{\text{eq}}(\phi') = \frac{1}{2\phi_m}, \qquad \phi' \in [-\phi_m, \phi_m].
\end{equation}
\item The \textbf{conditional (propagator)} probability $P(\phi, t \,|\, \phi')$: the probability of being at angle $\phi=\phi(t'+t)$ after a lag time $t$, given that the particle was at $\phi'$ at the reference time $t'$.
\end{enumerate}

The propagator for diffusion on $[-\phi_m, \phi_m]$ with reflecting boundary conditions is:
\begin{align}
P(\phi,t \mid \phi') &=
\frac{1}{2\phi_m}
\bigg[
1
+2\sum_{n=1}^{\infty}
e^{-D_r\left(\frac{n\pi}{\phi_m}\right)^2t}
\cos\!\left(\frac{n\pi\phi}{\phi_m}\right)
\cos\!\left(\frac{n\pi\phi'}{\phi_m}\right) \notag \\
& +2\sum_{n=1}^{\infty}
e^{-D_r\left(\frac{(2n-1)\pi}{2\phi_m}\right)^2t}
\sin\!\left(\frac{(2n-1)\pi\phi}{2\phi_m}\right)\notag \\
& \sin\!\left(\frac{(2n-1)\pi\phi'}{2\phi_m}\right)
\bigg].
\end{align}

\begin{widetext}
The autocorrelation becomes
\begin{equation}
C^{\text{*}}(t) = \int_{-\phi_m}^{\phi_m} d\phi' \; P_{\text{eq}}(\phi')
\int_{-\phi_m}^{\phi_m} d\phi \; \cos(\phi - \phi') \; P(\phi, t \,|\, \phi').
\label{eq:slide-def}
\end{equation}
Since
\begin{equation}
\cos(\phi - \phi') = \cos\phi\,\cos\phi' + \sin\phi\,\sin\phi',
\label{eq:cos-expand}
\end{equation}
we get
\begin{equation}
C^{\text{*}}(t)
=
\underbrace{
\int_{-\phi_m}^{\phi_m} d\phi'\, P_{\text{eq}}(\phi')
\int_{-\phi_m}^{\phi_m} d\phi\,
\cos\phi\,\cos\phi'\,
P(\phi,t|\phi')
}_{I_{\cos}}
+
\underbrace{
\int_{-\phi_m}^{\phi_m} d\phi'\, P_{\text{eq}}(\phi')
\int_{-\phi_m}^{\phi_m} d\phi\,
\sin\phi\,\sin\phi'\,
P(\phi,t|\phi')
}_{I_{\sin}}.
\label{eq:split}
\end{equation}
\end{widetext}

\begin{widetext}
Evaluation of the inner integrals. We first compute the inner $\phi$ integrals. Consider
\begin{equation}
\mathcal{I}_c(t, \phi') \equiv \int_{-\phi_m}^{\phi_m} \cos\phi \; P(\phi, t|\phi') \; d\phi .
\label{eq:Ic-def}
\end{equation}
By parity, $\cos\phi$ is even and the odd (sine) eigenfunctions of the propagator are odd, so
\begin{equation}
\int_{-\phi_m}^{\phi_m}\cos\phi\,
\sin\!\left(\frac{(2n-1)\pi\phi}{2\phi_m}\right)d\phi = 0 ,
\qquad
\int_{-\phi_m}^{\phi_m}\sin\phi\,
\cos\!\left(\frac{n\pi\phi}{\phi_m}\right)d\phi = 0 .
\label{eq:parity}
\end{equation}
Only the even (cosine) modes therefore contribute to $\mathcal{I}_c$, and only the odd
(sine) modes to $\mathcal{I}_s$ below. Substituting the propagator,
\begin{equation}
\mathcal{I}_c(t,\phi') = \frac{1}{2\phi_m}\int_{-\phi_m}^{\phi_m} \cos\phi
\left[1 + 2\sum_{n=1}^{\infty} e^{-(n\pi/\phi_m)^2 D_r t}
\cos\!\left(\frac{n\pi\phi}{\phi_m}\right)\cos\!\left(\frac{n\pi\phi'}{\phi_m}\right)\right] d\phi .
\label{eq:Ic-sub}
\end{equation}

The required integrals are
\begin{equation}
\int_{-\phi_m}^{\phi_m} \cos\phi \; d\phi = 2\sin\phi_m ,
\qquad
\int_{-\phi_m}^{\phi_m} \cos\phi \; \cos\!\left(\frac{n\pi\phi}{\phi_m}\right) d\phi
= \frac{2(-1)^n\,\sin\phi_m}{1 - n^2\pi^2/\phi_m^2}
= \frac{2(-1)^n\,\phi_m^2\,\sin\phi_m}{\phi_m^2 - n^2\pi^2} ,
\label{eq:orth-cos}
\end{equation}
which follow from $\int_{-a}^{a}\cos(\alpha x)\cos(n\pi x/a)\,dx$ with $\alpha=1$ and
$a=\phi_m$. Defining
\begin{equation}
A_n \equiv \frac{(-1)^n\,\sin\phi_m}{\phi_m^2 - n^2\pi^2} ,
\label{eq:An}
\end{equation}
Eq.~\eqref{eq:Ic-sub} becomes
\begin{equation}
\mathcal{I}_c(t,\phi') = \frac{\sin\phi_m}{\phi_m}
+ \frac{1}{\phi_m}\sum_{n=1}^{\infty} e^{-(n\pi/\phi_m)^2 D_rt}\;
\underbrace{\frac{2(-1)^n\,\phi_m^2\,\sin\phi_m}{\phi_m^2 - n^2\pi^2}}_{=\,2\phi_m^2 A_n}\;
\cos\!\left(\frac{n\pi\phi'}{\phi_m}\right) ,
\label{eq:Ic-mid}
\end{equation}
that is,
\begin{equation}
\mathcal{I}_c(t,\phi') = \frac{\sin\phi_m}{\phi_m}
+ 2\phi_m\sum_{n=1}^{\infty} A_n\; e^{-(n\pi/\phi_m)^2 D_rt}\;
\cos\!\left(\frac{n\pi\phi'}{\phi_m}\right) .
\label{eq:Ic-final}
\end{equation}

Outer integral $I_{\cos}$

With $P_{\rm eq}(\phi')=1/2\phi_m$,
\begin{equation}
I_{\cos} = \frac{1}{2\phi_m}\int_{-\phi_m}^{\phi_m}\cos\phi'\;\mathcal{I}_c(t,\phi')\;d\phi'
= \frac{1}{2\phi_m}\int_{-\phi_m}^{\phi_m}\cos\phi'
\left[\frac{\sin\phi_m}{\phi_m} + 2\phi_m\sum_{n=1}^{\infty}A_n\,
e^{-(n\pi/\phi_m)^2 D_rt}\cos\!\left(\frac{n\pi\phi'}{\phi_m}\right)\right]d\phi' .
\label{eq:Icos-sub}
\end{equation}
The first term gives $\frac{1}{2\phi_m}\cdot\frac{\sin\phi_m}{\phi_m}\cdot 2\sin\phi_m
= \sin^2\phi_m/\phi_m^2$. For the second, Eq.~\eqref{eq:orth-cos} gives
\begin{align}
\frac{1}{2\phi_m}\cdot 2\phi_m\sum_{n=1}^{\infty}A_n\,e^{-(n\pi/\phi_m)^2 D_rt}
\int_{-\phi_m}^{\phi_m}\cos\phi'\cos\!\left(\frac{n\pi\phi'}{\phi_m}\right)d\phi'
&= \sum_{n=1}^{\infty}A_n\,e^{-(n\pi/\phi_m)^2 D_rt}\,
\frac{2(-1)^n\,\phi_m^2\,\sin\phi_m}{\phi_m^2-n^2\pi^2}
\nonumber\\[1mm]
&= 2\phi_m^2\sum_{n=1}^{\infty}A_n^2\;e^{-(n\pi/\phi_m)^2 D_rt} ,
\label{eq:Icos-sum}
\end{align}
where
\begin{equation}
A_n \cdot \frac{2(-1)^n\,\phi_m^2\,\sin\phi_m}{\phi_m^2 - n^2\pi^2}
= \frac{(-1)^n\sin\phi_m}{\phi_m^2-n^2\pi^2}\cdot
\frac{2(-1)^n\phi_m^2\sin\phi_m}{\phi_m^2-n^2\pi^2}
= \frac{2\phi_m^2\sin^2\phi_m}{(\phi_m^2-n^2\pi^2)^2} .
\label{eq:An-squared}
\end{equation}
Hence
\begin{equation}
I_{\cos} = \frac{\sin^2\phi_m}{\phi_m^2}
+ \frac{2\sin^2\phi_m}{\phi_m^2}\sum_{n=1}^{\infty}
\frac{e^{-D_r\left(\frac{n\pi}{\phi_m}\right)^2t}}
{\left[1-\left(\frac{n\pi}{\phi_m}\right)^2\right]^2} .
\label{eq:Icos-final}
\end{equation}

Sine contribution $I_{\sin}$

Define
\begin{equation}
\mathcal{I}_s(t,\phi') \equiv
\int_{-\phi_m}^{\phi_m} \sin\phi \; P(\phi,t\mid\phi')\,d\phi .
\label{eq:Is-def}
\end{equation}
The uniform term drops out since $\int_{-\phi_m}^{\phi_m}\sin\phi\,d\phi=0$, and by
Eq.~\eqref{eq:parity} only the odd eigenfunctions survive; this contribution is therefore
nonzero. The relevant integral is
\begin{equation}
\int_{-\phi_m}^{\phi_m}\sin\phi\,
\sin\!\left(\frac{(2n-1)\pi\phi}{2\phi_m}\right)d\phi
= \frac{2(-1)^{n}\,\cos\phi_m}{1-\left(\frac{(2n-1)\pi}{2\phi_m}\right)^{2}}
= \frac{2(-1)^{n}\,\phi_m^{2}\,\cos\phi_m}
{\phi_m^{2}-\left(\frac{(2n-1)\pi}{2}\right)^{2}} .
\label{eq:orth-sin}
\end{equation}
In analogy with Eq.~\eqref{eq:An} we define
\begin{equation}
B_n \equiv \frac{(-1)^{n}\,\cos\phi_m}
{\phi_m^{2}-\left(\frac{(2n-1)\pi}{2}\right)^{2}} ,
\label{eq:Bn}
\end{equation}
so that
\begin{equation}
\mathcal{I}_s(t,\phi') = 2\phi_m\sum_{n=1}^{\infty} B_n\,
e^{-D_r\left(\frac{(2n-1)\pi}{2\phi_m}\right)^2t}
\sin\!\left(\frac{(2n-1)\pi\phi'}{2\phi_m}\right) .
\label{eq:Is-final}
\end{equation}
Averaging over the uniform equilibrium distribution and using Eq.~\eqref{eq:orth-sin} once more,
\begin{equation}
I_{\sin} = \frac{1}{2\phi_m}\int_{-\phi_m}^{\phi_m}\sin\phi'\,\mathcal{I}_s(t,\phi')\,d\phi'
= 2\phi_m^2\sum_{n=1}^{\infty}B_n^2\,
e^{-D_r\left(\frac{(2n-1)\pi}{2\phi_m}\right)^2t}
= \frac{2\cos^2\phi_m}{\phi_m^2}\sum_{n=1}^{\infty}
\frac{e^{-D_r\left(\frac{(2n-1)\pi}{2\phi_m}\right)^2t}}
{\left[1-\left(\frac{(2n-1)\pi}{2\phi_m}\right)^2\right]^2} .
\label{eq:Isin-final}
\end{equation}
combine all the term $I_{\cos}$ and $I_{\sin}$ an we get:
\begin{equation}
C(t)
=
\frac{\sin^2\phi_m}{\phi_m^2}
+
\frac{2\sin^2\phi_m}{\phi_m^2}
\sum_{n=1}^{\infty}
\frac{
e^{-D_r\left(\frac{n\pi}{\phi_m}\right)^2t}
}{
\left[1-\left(\frac{n\pi}{\phi_m}\right)^2\right]^2
}
+
\frac{2\cos^2\phi_m}{\phi_m^2}
\sum_{n=1}^{\infty}
\frac{
e^{-D_r\left(\frac{(2n-1)\pi}{2\phi_m}\right)^2t}
}{
\left[1-\left(\frac{(2n-1)\pi}{2\phi_m}\right)^2\right]^2
}
\end{equation}

\end{widetext}

\clearpage
\twocolumngrid
\section{APPENDIX B: Entropy production rate of a communicating pair}
\label{app:epr}
Consider a pair of birds $i$ and $j$, and write
$a\equiv\mathcal{F}_{i\rightarrow j}$ and $b\equiv\mathcal{F}_{j\rightarrow i}$
for the rates of IC events in the two directions, taken as constant over the
flight. As a minimal model, we treat the two directed event streams as
independent Poisson processes: in any short interval $dt$, an event from $i$ to
$j$ occurs with probability $a\,dt$, an event from $j$ to $i$ with probability
$b\,dt$, and no event with probability $1-(a+b)\,dt$, independently of all other intervals. The probability that no event occurs during a total time $T$ is
therefore $\lim_{dt\to0}[1-(a+b)\,dt]^{T/dt}=e^{-(a+b)T}$. Hence over the total time $T$ containing $n_{\rightarrow}$ events from $i$ to $j$ and
$n_{\leftarrow}$ events from $j$ to $i$, at specified times and at no other time,
has probability density
\begin{equation}
\mathcal P^+=a^{n_{\rightarrow}}\,b^{n_{\leftarrow}}\,e^{-(a+b)T}.
\end{equation}
Its time-reversed probability density is obtained by interchanging the direction of
every event while keeping the event times [Fig.~\ref{fig4}(b)]. Its
$n_{\rightarrow}$ events must then be generated at rate $b$ and its
$n_{\leftarrow}$ events at rate $a$:
\begin{equation}
\mathcal P^-=a^{n_{\leftarrow}}\,b^{n_{\rightarrow}}\,e^{-(a+b)T}.
\end{equation}
The exponential factors cancel in the ratio, giving
\begin{equation}
\ln\frac{\mathcal P^+}{\mathcal P^-}=(n_{\rightarrow}-n_{\leftarrow})\ln\frac{a}{b}.
\end{equation}

Averaging over the forward process, for which
$\langle n_{\rightarrow}\rangle=aT$ and $\langle n_{\leftarrow}\rangle=bT$, and
dividing by $T$ yields the entropy production rate,
i.e.\ the Kullback-Leibler divergence rate between forward and direction-reversed distributions, in units of $k_B$:
\begin{equation}
\dot S_{ij}=\frac{1}{T}\Bigl\langle\ln\frac{\mathcal P^+}{\mathcal P^-}\Bigr\rangle
=(a-b)\ln\frac{a}{b}\;\geq\;0 .
\end{equation}
It vanishes only in the case of reciprocal communication, where $a=b$.

\section{APPENDIX C:
Analytical evaluation of the pairwise sum}

We consider the following pairwise sum shown in Eq. \ref{eq:initial_sum}:

\begin{equation}
\dot{S} = \frac{\mathcal R_0}{2}
\sum_{i=1}^{N}
\sum_{\substack{j=1\\j\neq i}}^{N}
\eta(j-i)
\left[
e^{-\eta(i-1)}
-
e^{-\eta(j-1)}
\right].
\label{eq:initial_sum1}
\end{equation}

For convenience, we define
\begin{equation}
q=e^{-\eta}.
\label{eq:q_definition}
\end{equation}
Then Eq.~\eqref{eq:initial_sum} can be written as
\begin{equation}
\dot{S} =
\frac{\eta \mathcal R_0}{2}
\sum_{i=1}^{N}
\sum_{\substack{j=1\\j\neq i}}^{N}
(j-i)
\left(q^{i-1}-q^{j-1}\right).
\label{eq:q_sum}
\end{equation}

The contribution from the pair $(i,j)$ is identical to that from
the pair $(j,i)$. Therefore,
\begin{equation}
\dot{S} = \eta \mathcal R_0
\sum_{i<j}
(j-i)
\left(q^{i-1}-q^{j-1}\right).
\label{eq:symmetric_sum}
\end{equation}

We now collect the coefficient of $q^{k-1}$. For a given $k$, this
coefficient is
\begin{align}
&\sum_{j=k+1}^{N}(j-k)
-
\sum_{i=1}^{k-1}(k-i)
\nonumber\\
&=
\frac{(N-k)(N-k+1)}{2}
-
\frac{k(k-1)}{2}
\nonumber\\
&=
\frac{N(N+1-2k)}{2}.
\label{eq:coefficient}
\end{align}

Substituting Eq.~\eqref{eq:coefficient} into Eq.~\eqref{eq:symmetric_sum},
we obtain
\begin{equation}
\dot{S} =
\frac{N\eta \mathcal R_0}{2}
\sum_{k=1}^{N}
(N+1-2k)q^{k-1}.
\label{eq:reduced_sum}
\end{equation}

Introducing $m=k-1$, Eq.~\eqref{eq:reduced_sum} becomes
\begin{equation}
\dot{S} =
\frac{N\eta\mathcal R_0}{2}
\sum_{m=0}^{N-1}
(N-1-2m)q^m.
\label{eq:m_sum}
\end{equation}

The required finite geometric sums are
\begin{equation}
\sum_{m=0}^{N-1}q^m
=
\frac{1-q^N}{1-q},
\label{eq:geometric_sum}
\end{equation}
and
\begin{equation}
\sum_{m=0}^{N-1}mq^m
=
\frac{
q-Nq^N+(N-1)q^{N+1}
}{
(1-q)^2
}.
\label{eq:weighted_geometric_sum}
\end{equation}

Using Eqs.~\eqref{eq:geometric_sum} and
\eqref{eq:weighted_geometric_sum} in Eq.~\eqref{eq:m_sum}, we obtain
\begin{align}
\dot{S}
&=
\frac{N\eta\mathcal R_0}{2}
\left[
(N-1)\frac{1-q^N}{1-q}
-
2\frac{
q-Nq^N+(N-1)q^{N+1}
}{
(1-q)^2
}
\right].
\label{eq:before_final}
\end{align}

Putting the terms over a common denominator gives
\begin{align}
\dot{S}
&=
\frac{N\eta\mathcal R_0}{2(1-q)^2}
\Big[
(N-1)
-(N+1)q
\nonumber\\
&\qquad\qquad
+(N+1)q^N
-(N-1)q^{N+1}
\Big].
\label{eq:q_final}
\end{align}

Finally, substituting $q=e^{-\eta}$, we obtain
\begin{equation}
\begin{aligned}
\dot{S} &=
\frac{N\eta\mathcal R_0}{2(1-e^{-\eta})^2}
\Big[
(N-1)-(N+1)e^{-\eta} \\
&\qquad +(N+1)e^{-N\eta}
-(N-1)e^{-(N+1)\eta}
\Big].
\end{aligned}
\label{eq:final_result_A}
\end{equation}


\clearpage
\onecolumngrid

\setcounter{equation}{0}
\setcounter{figure}{0}
\setcounter{table}{0}
\setcounter{section}{0}

\renewcommand{\theequation}{S\arabic{equation}}
\renewcommand{\thefigure}{S\arabic{figure}}
\renewcommand{\thetable}{S\arabic{table}}

\begin{center}
{\bfseries
Supplemental Material for:\\
Entropy-production cost limits hierarchical decision-making in homing pigeon flocks
\par}
\vspace{1.5em}

Bibaswan Ghosh$^{1,*}$, Vipul Rai$^{2,*}$,
Harsh Soni$^{2}$, and Nitin Kumar$^{1}$
\par
\vspace{0.8em}

{\small\itshape
$^{1}$Department of Physics, Indian Institute of Technology Bombay,
Powai, Mumbai 400076, India\\
$^{2}$School of Physical Sciences, Indian Institute of Technology Mandi,
Mandi 175005, India
\par}
\vspace{0.5em}

{\small $^*$Equal contribution\\
nkumar@iitb.ac.in}
\end{center}

\vspace{1.5em}

\section{I. Validating the homing model on different animal species}

We calculate the temporal orientation autocorrelation directly from two-dimensional GPS trajectory data. $\hat{\mathbf{n}}(t)$ is defined as the instantaneous unit displacement vector between two successive frames,
\begin{equation}
\hat{\mathbf{n}}(t)=
\frac{\mathbf{r}(t+\Delta t)-\mathbf{r}(t)}
{|\mathbf{r}(t+\Delta t)-\mathbf{r}(t)|},
\end{equation}
where $\mathbf{r}(t)=(x(t),y(t))$ and $\Delta t$ is the sampling interval.
The temporal orientation autocorrelation function is then calculated as
\begin{equation}
C(t)
=
\left\langle
\hat{\mathbf{n}}(t'+t)\cdot\hat{\mathbf{n}}(t')
\right\rangle_{t'},
\end{equation}
where $t$ is the time lag. The subscript $t'$ indicates a sliding
average over all available time origins along the trajectory.

\begin{figure*}[h]
	\includegraphics[width= 0.6 \textwidth]{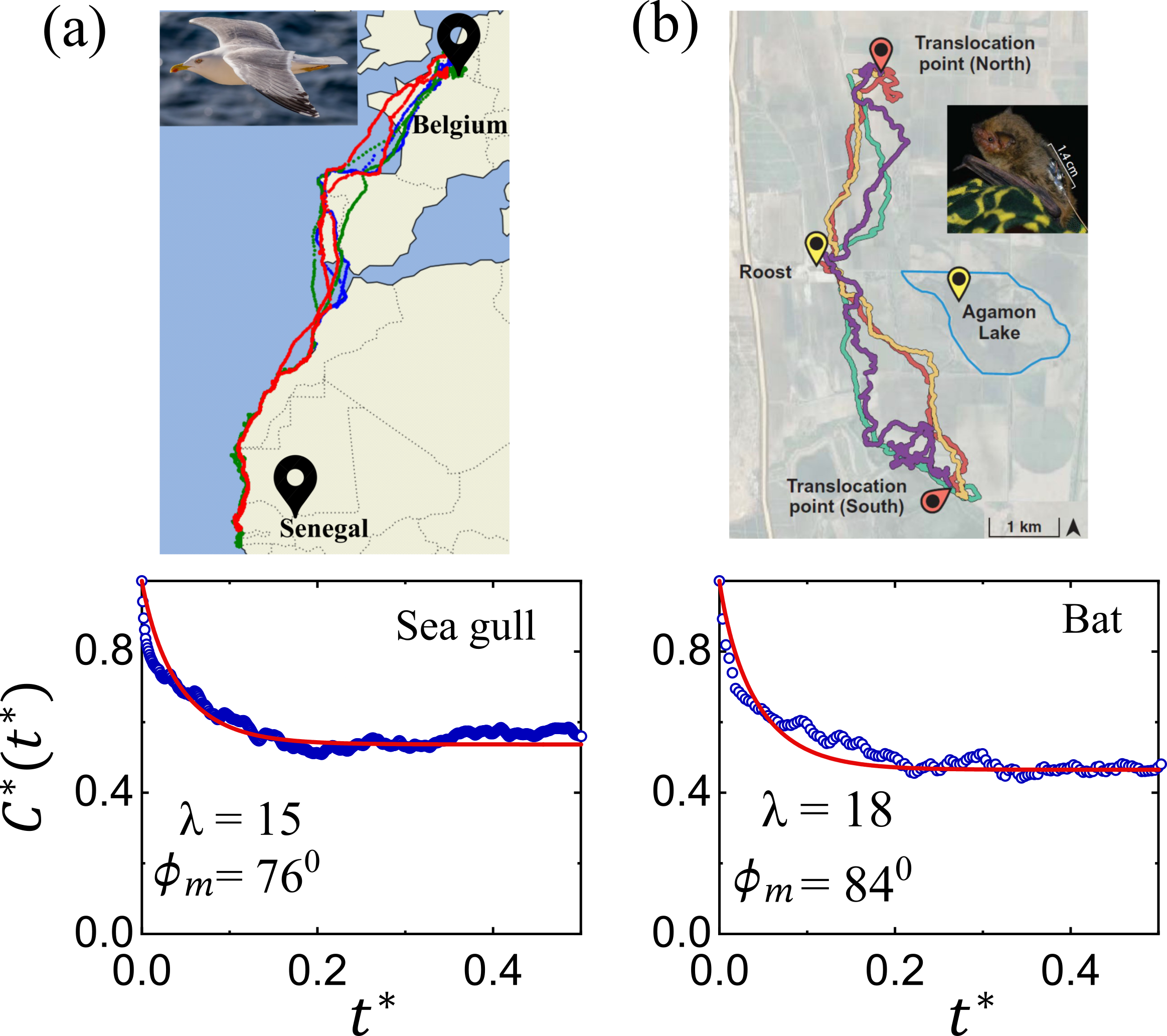}
	\caption{Validation of our homing model by fitting Eq. 1 of the main paper to trajectories of different animal species. The images in the upper row are adapted from \cite{seegul_website} and \cite{goldshtein2024acoustic} for (a) and (b) respectively.}
	\label{Sap_Fig1}
\end{figure*}

Within our homing model framework, the navigational dynamics of different species can be described using a single theoretical formalism. As shown in Fig.[\ref{Sap_Fig1}], the trajectories of seagulls \cite{seegul_website} and bats \cite{goldshtein2024acoustic} span a wide range of space and time scales, distinct movement patterns, ecological contexts, and sensory capabilities. Nevertheless, the $C^*(t^*)$ calculated from these trajectories is accurately captured by Eq. 1 of the main paper. 

From the dataset of previously published papers, we extracted the trajectories and calculated the orientation autocorrelation function directly from the recorded flight paths. We then fitted the resulting autocorrelation functions using our homing model. We used a constant speed and a release-to-home distance specific to each dataset. For the bat trajectories, we used a constant flight velocity of ($v = \text{20 km/h}$) and a release-to-home distance of ($L_H = \text{3}$ km). For the seagull trajectories, we used ($v = \text{36 km/h}$) and ($L_H = \text{5000 km}$), based on the experimental conditions. These species-specific values were kept fixed during the fitting procedure, while the remaining parameters ($\lambda$ and $\phi_m$) were obtained from the fit.

\section{II. Applying homing model to different flock sizes of pigeons}
To ensure a consistent comparison across flock sizes, we used the same $v$ (= 55 km/h) and $L_H$ (= 8.1 km) for each dataset. Before calculating $C(t)$ for the data of reference \cite{sasaki2022empirical}, we excluded a few trials in which the trajectories contained substantial tracking or flight anomalies. This includes looping (flying in circles) during flight and incomplete flight paths. For the solo flights, trials 11, 23, 27, 30, 41, 49, 57, 59, 86, 93, 116, 121, 138   were excluded because the birds exhibited pronounced looping behavior. For flock size 2, trials 3, 7, and 11 were excluded as the trajectory of one bird terminated substantially earlier than the others. Similarly, for flock size 8, trials 2 and 3 were excluded as several birds' flights ended well before the arrival to loft. The remaining trajectories were then used to calculate the $C(t)$ and fit the resulting curves using the homing equation (Eq. 1 of the main paper), with the same values of \(v\) and \(L_H\) maintained across all flock sizes.

\section{III. Simulation details}
\subsection{A. ABP homing simulation details}
To reproduce empirical homing trajectories numerically, we model each individual as an active Brownian particle (ABP) moving at a constant self-propulsion speed (Fig. 1a of the main paper). The position \(\mathbf{r}=(x,y)\) and orientation \(\theta\) (measured with respect to the positive $x$ axis) of the ABP evolve according to

$$
\frac{d\mathbf{r}}{dt}
=
v
\begin{pmatrix}
\cos\theta\\
\sin\theta
\end{pmatrix},
$$

and

$$
\frac{d\theta}{dt}
=
\sqrt{2D_r}\,\chi(t),
$$

where $v$ is the constant self-propulsion speed, $D_r$ is the rotational diffusion coefficient, and $\chi(t)$ is a Gaussian white-noise term with zero mean and unit variance, satisfying

$$
\langle \chi (t)\rangle=0,
\qquad
\langle\chi(t)\chi(t')\rangle
=
\delta(t-t').
$$

The equations are integrated numerically using a time step of $\Delta t=0.2$ s, which matches the temporal resolution of the experimentally recorded trajectories.

\begin{figure*}
	\includegraphics[width= 0.6 \textwidth]{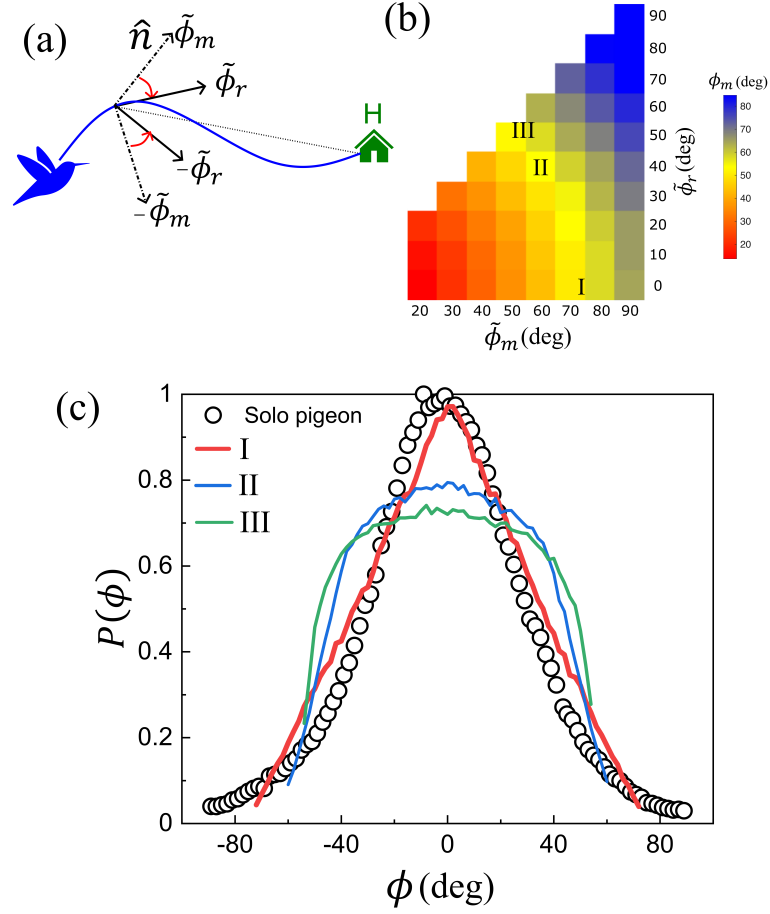}
	\caption{\textbf{(a)} \textbf{Schematic of the resetting mechanism}: When an ABP deviates from the home direction by $\pm\tilde{\phi}_m$, its orientation is reset to $\pm\tilde{\phi}_r$, respectively, and it resumes its motion toward home. \textbf{(b)} Heat map of the combinations of $\tilde{\phi}_m$ and $\tilde{\phi}_r$ that yield the target value $\phi_m=54^\circ$. \textbf{(c)} Orientation distributions $P(\phi)$ for different parameter combinations yielding the same $\phi_m$. The combination $(\tilde{\phi}_m,\tilde{\phi}_r)=(74^\circ,0^\circ)$ provides the closest match to the empirical target distribution for solo pigeons. Parameters I, II, and III are marked in (b) are ($\tilde\phi_m, \tilde\phi_r$) = (74$^\circ$, 0$^\circ$), (60$^\circ$, 40$^\circ$), and (54$^\circ$,53$^\circ$) respectively.}
	\label{Sap_Fig2}
\end{figure*}

Starting from ABPs, we introduce a resetting protocol to model directed navigation toward the home location. For each simulation, the particles are initialized with orientations perfectly aligned with the homeward direction ($\phi = 0^\circ$), and their orientations are updated at each time step according to stochastic rotational-diffusion dynamics. We use $D_r = 0.02$ rad$^2$s$^{-1}$, which gives us $\lambda=10$. When an agent's heading direction ($\phi$) deviates from the beeline path to $H$ by more than a threshold angle, $\pm\tilde{\phi}_m$, its trajectory is corrected by a finite angular displacement, such that its new heading direction is set to $\pm\tilde{\phi}_r$ [Fig.~\ref{Sap_Fig2}\textbf{a}] with respect to the home direction. In simulations, we use a constant speed of $v=55$ km h$^{-1}$, which is the same as that used when fitting the experimental dataset. The characteristic release-to-home distance is taken as ($L_H=8.1$ km). 

\subsection{B. Identifying the Resetting Protocol from the \texorpdfstring{$\boldsymbol{\phi}$}{phi} Distribution}

\begin{figure*}
	\includegraphics[width= 0.6 \textwidth]{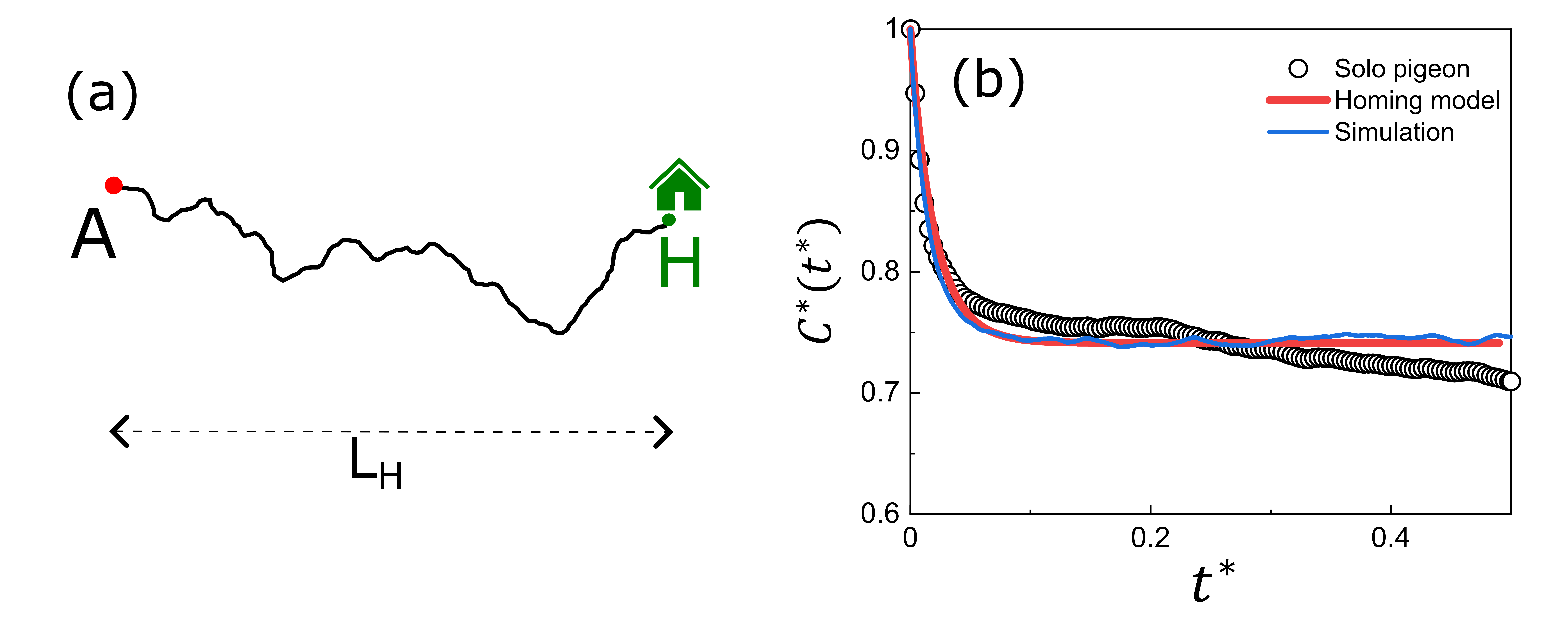}
	\caption{\textbf{(a)} Typical homing trajectory of simulated ABP particle with parameters ($\tilde\phi_m, \tilde\phi_r$) = (74$^\circ$, 0$^\circ$). The resulting $C^*(t^*)$ for simulation (thinner blue line) shows an excellent agreement with the fitted Eq. 1 from the main paper (thicker pink line) to the empirical function (circle symbol).}
	\label{Sap_Fig3}
\end{figure*}

By fitting the trajectories of solo pigeons\cite{sasaki2022empirical} using the homing equation (Eq. 1 of the main paper), we obtain a characteristic angle of $\phi_m = 54^\circ$. Different combinations of $\tilde{\phi}_m$ and $\tilde{\phi}_r$ can produce the same value of $\phi_m = 54^\circ$. Therefore, matching the value of $\phi_m$ alone does not uniquely determine the correct resetting protocol. To select the appropriate combination, we further compare the resulting distribution of $\phi$ with that observed in the pigeon trajectories. We choose the combination of $\tilde{\phi}_m$ and $\tilde{\phi}_r$ that not only reproduces the fitted value of $\phi_m$, but also has an orientation distribution similar to that of the pigeons. In Fig.~\ref{Sap_Fig3}, we show that the orientation autocorrelation of our simulated trajectories exhibits the same statistical features as that observed for solo pigeons.

\section{IV. Self-correction rate}
\begin{figure}[h]
	\includegraphics[width= 0.7 \textwidth]{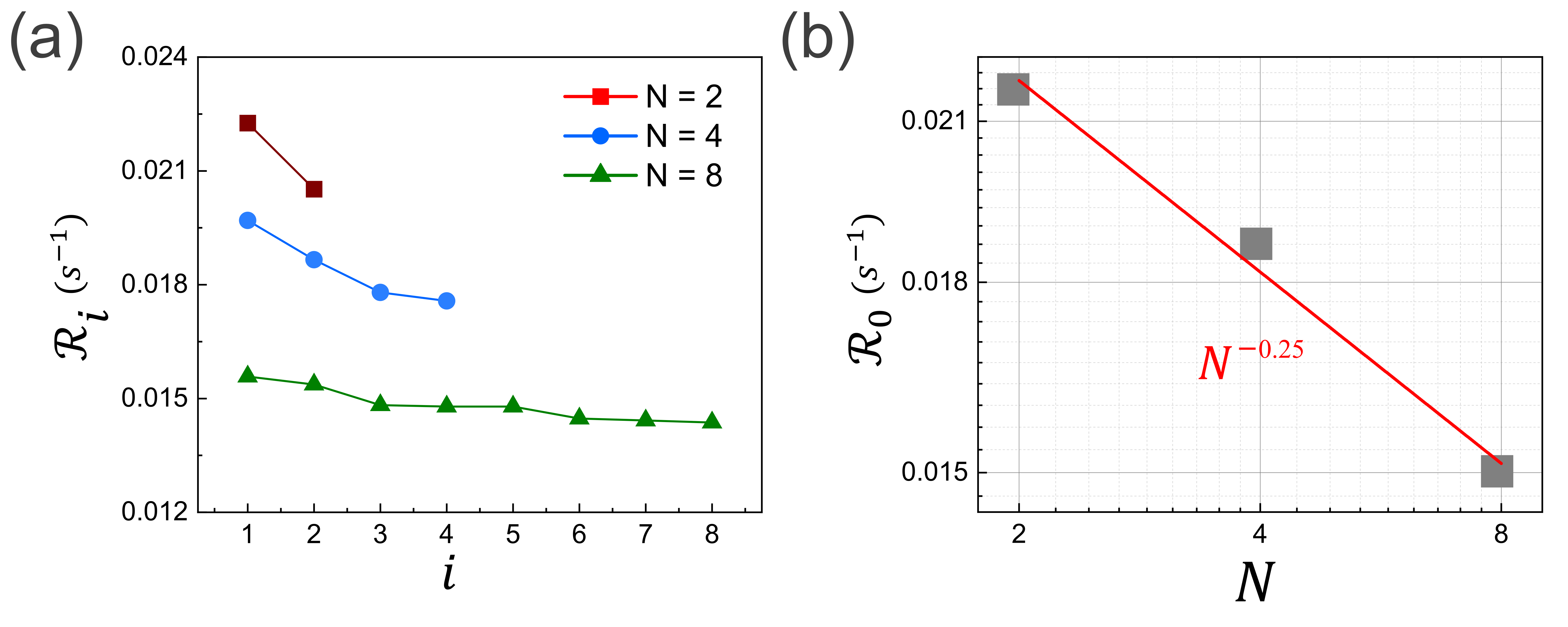}
	\caption{\textbf{(a)} The mean of the number of self-correction events per unit time ($\mathcal R_i$) over a complete homing event for different flock sizes, $N$. \textbf{(b)} The average of $\mathcal R_i$ over all birds ($\mathcal R_0$) in a given flock size decreases with its size $N$ as $N^{-0.25}$.}
	\label{Sap_Fig4}
\end{figure}

For each flock size, we find that the rate of SC events, $\mathcal{R}_i$, defined as the total number of events divided by its homing time and averaged over multiple iterations, remains largely constant across all agents within the flock (See Fig. \ref{Sap_Fig4}\textbf{a}). Additionally, we find that the mean SC event rate, $\mathcal{R}_0$, obtained by averaging $\mathcal{R}_i$ over all agents in the flock, scales as $\mathcal{R}_0 =0.026 N^{-0.25}$ [solid line in Fig. \ref{Sap_Fig4}\textbf{b}].

\section{V. Delay time calculation}

For each trial containing \(N\) birds, we extracted the \(x\)- and \(y\)-coordinates of each bird's trajectory.

The instantaneous displacement vector of bird \(i\) was calculated as
\begin{equation}
\Delta \mathbf{r}_i(t)
=
\mathbf{r}_i(t+\Delta t)-\mathbf{r}_i(t),
\end{equation}
where \(\Delta t\) is the sampling interval. The displacement vectors were normalized to obtain the instantaneous direction of motion,
\begin{equation}
\hat{\mathbf{v}}_i(t)
=
\frac{\Delta \mathbf{r}_i(t)}
{|\Delta \mathbf{r}_i(t)|}.
\end{equation}

For every pair of distinct birds \(i\) and \(j\), we calculate the temporal cross-correlation of their velocity directions over a range of positive and negative time lags,
\begin{equation}
k_{ij}(\tau)
=
\left\langle
\hat{\mathbf{v}}_i(t)
\cdot
\hat{\mathbf{v}}_j(t+\tau)
\right\rangle ,
\end{equation}
where \(\langle\cdot\rangle_t\) denotes an average over all time points for which both trajectories are available. Positive \(\tau\) corresponds to the motion of bird \(j\) being compared with the subsequent motion of bird \(i\), whereas negative \(\tau\) corresponds to the opposite temporal ordering.

The correlation was evaluated for time lags
\begin{equation}
\tau = -\tau_{\max},\ldots,0,\ldots,+\tau_{\max},
\end{equation}
with \(\tau_{\max}=20\) sec. The lag corresponding to the maximum cross-correlation was identified as the characteristic delay between the two birds,
\begin{equation}
\tau_{ij}^{*}
=
\underset{\tau}{\operatorname{arg\,max}}\,
k_{ij}(\tau).
\end{equation}

To obtain a single hierarchy measure for each bird, we averaged its pairwise delay times relative to all other birds in the flock,
\begin{equation}
\tau^*_i
=
\frac{1}{N-1}
\sum_{\substack{j=1 \\ j\neq i}}^{N}\tau_{ij}^{*}.
\end{equation}
In the present analysis, pairwise delays up to $\pm 1.5 $ s were considered for averaging. In case no significant correlation was observed in the $ij$ pair, it was not considered for averaging.

Finally, birds were ranked according to \(\tau^*_i\) in descending order. The resulting quantity \(\tau^*_i\) therefore provides a global measure of the temporal position of bird \(i\) relative to the other members of the flock. Birds with larger values of \(\tau^*_i\) are assigned higher positions in the resulting hierarchy.

\begin{figure*}
	\includegraphics[width= 0.4 \textwidth]{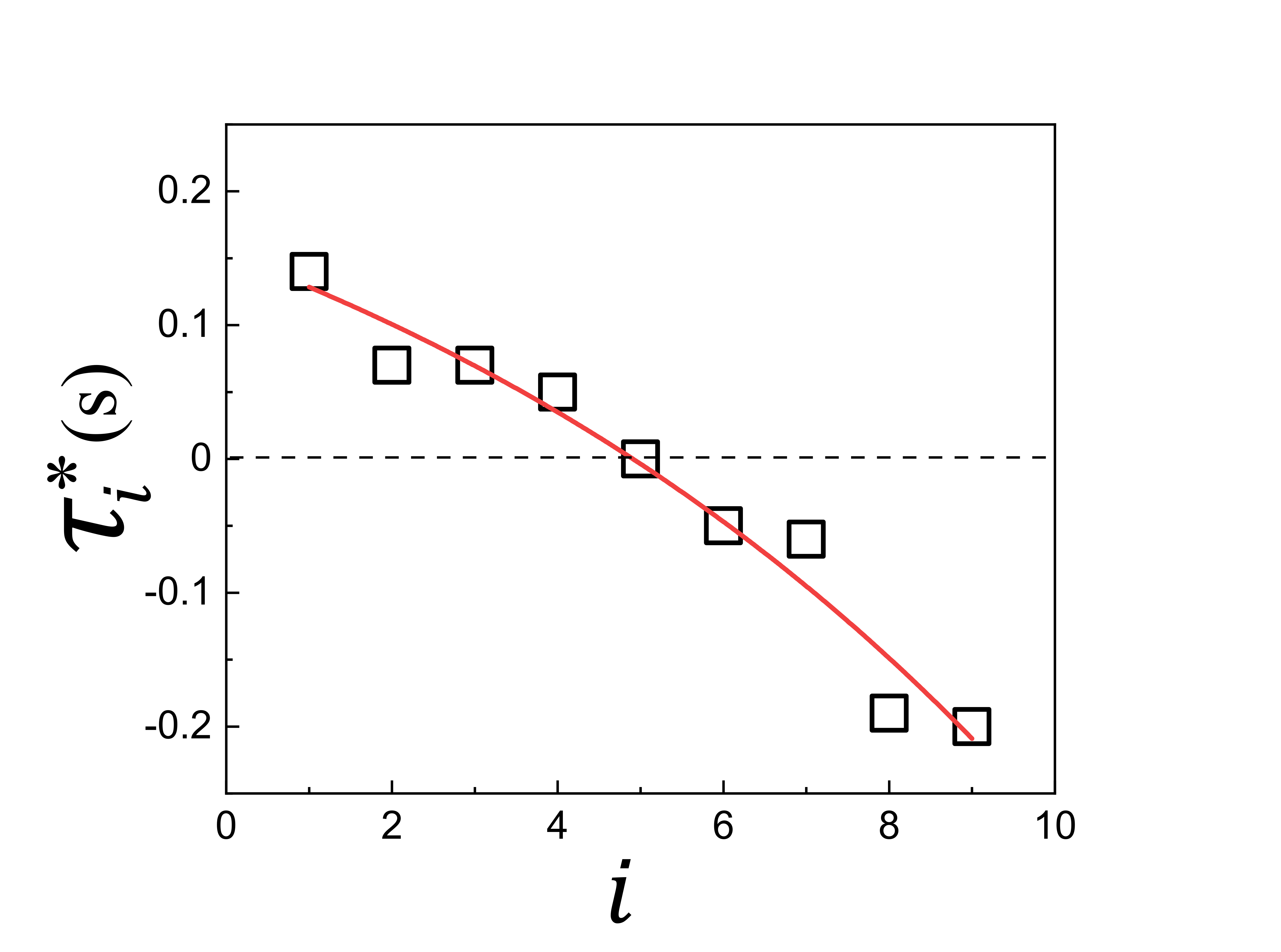}
	\caption{Fit of data provided in Fig. 3(a) of Nagy et al. \cite{nagy2010hierarchical} with Eq. 2 of the main paper.}
	\label{Sap_Fig5}
\end{figure*}

\section{VI. Analyzing the data set of flock size 9}
From the publicly available dataset accompanying\cite{nagy2010hierarchical}, we obtained trajectory data for homing pigeons flying in flocks of size 9. From the available recordings, we considered only the homing flights. Later we calculate the delay times from the trajectory data following the procedure described in Sec. V of the Supplementary Information. 

Their paper also reports the corresponding delay times which they calculated following their assumptions/protocols. Our calculated values show slight quantitative differences from the reported values, which may arise from minor differences in the analysis protocol used to extract the delay times. Nevertheless, our results reproduce the overall trend and fall within a similar range as those reported in the original study.
In Fig. \ref{Sap_Fig5}, we apply Eq. (2) of the main paper to the data provided in Fig. 3(a) of Nagy et al. \cite{nagy2010hierarchical}. The data are fitted using the parameters $\mathcal B = 0.24$ s, and $\eta = 0.11$. 


\clearpage

\renewcommand{\refname}{References}

\end{document}